\documentclass[a4paper,12pt]{article}
\usepackage{physics}
\usepackage{color}
\usepackage{amsmath, amssymb, amsfonts}
\usepackage{geometry}
\usepackage{graphicx}
\usepackage{titlesec}
\usepackage{pdfpages}
\usepackage{wrapfig}
\usepackage{lipsum}
\usepackage{url}
\usepackage{relsize}
\graphicspath{ {images/} }
\numberwithin{equation}{section} %Labels equations by section.

\usepackage{verbatim}
\usepackage{graphicx}
\usepackage{float}
\usepackage{color}

\definecolor{mygreen}{rgb}{0,0.6,0}
\definecolor{mygray}{rgb}{0.5,0.5,0.5}
\definecolor{mymauve}{rgb}{0.58,0,0.82}
\definecolor{light-gray}{gray}{0.95}

\begin{document}
	\title{The Superradiant Instability in AdS}
	\author{Joseph M.U. Sullivan}
	\date{May 5, 2016}
	\maketitle
			
	\begin{abstract}
		We consider the intermediate and end state behavior of the superradiantly perturbed Kerr black hole. Superradiant scattering in an asymptotically flat background is considered first. The case of a Kerr black hole in an Anti de-Sitter background is then discussed. Specifically we review what is known about the superradiant instability arising in AdS and its possible end state behavior.
	\end{abstract}
	
	\tableofcontents
\section*{Acknowledgements}
I would like to dedicate this essay to the memory of Professor Steve Detweiler. It was during our many conversations during my time as an undergraduate that he first spurred my interest in the physics of black holes. I am just one of many students who has benefited immensely from his guidance and insight. 

I would like to thank Dr. Jorge Santos for shepparding me toward a better understanding of the problems at the heart of this essay, for taking the time to thoroughly review \& critique my work and for the introduction to AdS. I also owe Dr. Mike Blake a debt of gratitude for the mathematical clarifications he provided. While I'm at it, I should probably also thank my mother.
\newpage
\section{Introduction}
This essay is concerned with the phenomena of superradiance in the setting of the Kerr black hole (Kerr BH). Superradiance is a wave phenomena in which an ingoing wave scatters off an object and in the process extracts some energy; the scattered wave is more energetic than the incident one. This is a particularly interesting process in the context of black hole physics because it provides an outlet for energy dissipation by a BH. Much of our discussion will be concerned with the results of \cite{Santos} and \cite{russ}. In \cite{russ} superradiant scattering was studied in the space-time  of a Kerr BH in Minkowski background. Frequency dependent conditions for superradiance and calculations of the extent of amplification were obtained; we will review these.  \bigskip \newline \indent
We also consider superradiance in a spacetime comprised of a Kerr BH in an Anti de-Sitter (AdS) background. While superradiant scattering in a Minkowski background is interesting in its own right, we will see that the corresponding problem in AdS is much richer and more complex. This is due in large part to the box-like nature of AdS. A scattered wave can now reflect off the boundary at infinity and return, in finite time, to the Kerr BH to extract more energy. This process can repeat many times suggesting that the Kerr-AdS BH is susceptible to a superradiant instability. This motivates a plethora of fascinating questions; can we characterize this instability, is there a relationship between general instabilities and these superradaint ones, what is the end state of the superradaintly perturbed Kerr-AdS BH...etc? In an effort to answer these questions we will draw heavily from the results of \cite{Santos}.
\bigskip
 Rather than dive right into a discussion of superradiance in the two Kerr BH spacetimes, we first provide an introduction to some of the objects and concepts fundamental to the problem. To begin we give a treatment of the Kerr BH. Next we discuss the ergoregion of the Kerr BH and the associated phenomena of the Penrose Process to motivate the idea of Kerr BHs being susceptible to energy extraction. To aid our discussion of the superradiant instability we give an overview of some of the important properties of AdS. We next introduce the Teukolsky formalism and comment on its importance to Kerr BH perturbation theory. Finally, with the Teukolsky formalism in our tool box, we give a generic analysis of BH superradiance. 

\subsection{The Kerr Black Hole}
A generic uncharged rotating BH in Minkowski background belongs to the Kerr family. Remarkably, this is just a two parameter family of solutions characterized by $M$ and $J$ which describe the mass and angular momentum of the BH. In natural units, with $c=1$ and $G =1$, the metric \cite{Kerr} is given by:
\begin{equation} \label{eq:eq1}
 ds^2 = -\frac{\Delta}{\rho^2}\bigg(dt-a\sin^2\theta d\phi\bigg)^2 +\frac{\rho^2}{\Delta}dr^2 + \rho^2 d\theta^2 \frac{\sin^2\theta}{\rho^2}\bigg(adt-(r^2+a^2)d\phi\bigg)^2  
\end{equation}
with $$\Delta = r^2 + a^2 -2Mr = (r-r_+)(r-r_-), ~~ \rho^2 = r^2 + a^2\cos^2\theta$$ where the mass of the spinning BH is given by $M$ and the angular momentum is given by $J=aM$. We see that the metric has a coordinate singularity at the roots of $\Delta$ with the larger root $r_+ = M+\sqrt{M^2 - a^2}$ determining the event horizon and the smaller root $r_- = M-\sqrt{M^2 -a^2}$ corresponding to a Cauchy horizon.
\subsection{Ergoregion and Energy Extraction}
Note that the spacetime has the Killing vector fields (KVFs) $$k^a=\bigg(\frac{\partial}{\partial t}\bigg)^a, ~~ m^a=\bigg(\frac{\partial}{\partial \phi}\bigg)^a .$$
Observe that $k^ak_a = g_{tt}(r,\theta)$ which, while monotone with respect to $r$ and negative for sufficiently large $r$, is not strictly negative in the region $r>r_+, ~~\theta \in [0,\pi]$. Thus finding the roots of $g_{tt}$ we see $k^a$ is timelike in the region $r> r_{erg} = M+\sqrt{M^2 -a^2\cos^2\theta}$, null at $r = r_{erg}$ and actually spacelike in the region $r_+ <r<r_{erg}$. This latter region defines the ergoregion of the Kerr BH.  
\begin{figure}[h!]
	\includegraphics[width = 14cm]{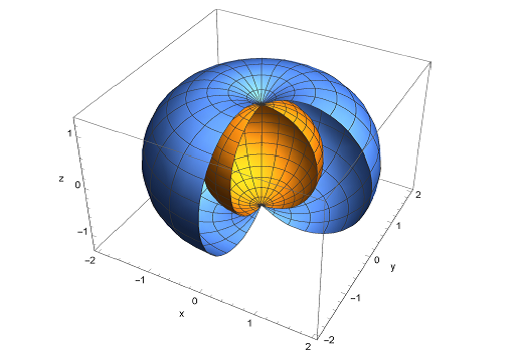}
	\caption{Plot from \cite{supover} of the ergoregion of a nearly extremal, $a\sim 0.999M$ Kerr BH. The ergoregion resides between the blue and orange surfaces. }
	\label{fig:ergo}
\end{figure}
\newline\indent 
A static observer, i.e person with 4-velocity parallel $ k^a$, is not allowed in the ergoregion as curves with tangent vector $\propto k^a$ are spacelike when $r<r_{ergo}$. Interpreting this physically, an observer cannot simply sit still in the ergoregion but is forced to rotate with the BH. 
\newline \indent 
On the other hand we can consider a stationary observer at constant $(r,\theta)$ with 4-velocity $$v^\mu = (\dot{t},0,0,\dot{\phi}) = \dot{t}(1,0,0,\Omega).$$ Such an observer can exist provided he/she travels on a timelike curve or equivalently $v^2<0$. This provides a condition for the existence of a stationary observer: $$v^2 \propto g_{tt} + 2\Omega g_{t\phi} + \Omega^2g_{\phi\phi} <0.$$ The zeros of the above expression are given by $$\Omega_{\pm} = \frac{-g_{t\phi} \pm \sqrt{g^2_{t\phi} - g_{tt}g_{\phi\phi}}}{g_{\phi\phi}} = \frac{-g_{t\phi}\pm \sqrt{\Delta}\sin\theta}{g_{\phi\phi}}$$ Note that $\Im({\Omega_\pm})\ne 0$ for $r_-<r<r_+$ hence there cannot be a stationary observer in this region. The permissible values are $\Omega \in [\Omega_-, \Omega_+]$. At $r_+,~ \Omega_-=\Omega_+$, meaning there is only one possibly stationary observer at the event horizon. \bigskip
\newline \indent
To motivate superradiance we now give an example of how a rotating BHs allow energy extraction. Suppose we have a particle with 4-momentum $P^a = \mu u^a$ which approaches the Kerr BH along a geodesic. The energy of the particle as measured by a static observer at infinity is conserved along the geodesic: $E =-k \cdot P$. Now suppose that inside the ergosphere the particle decays into two other particles with momenta $P_1^a ~~\&~~ P_2^a$. Momentum must be conserved so $P=P_1+P_2 \implies E = E_1 + E_2$. But since $k^a$ is spacelike in the ergoregion it is possible that $E_1 <0$ which implies that $E_2 = E-E_1>E$. It can be shown that $p_1$ must fall into the BH while $p_2$ can escape to infinity greater energy than the incident particle. Hence the BH will actually decrease in mass and energy will be extracted. \bigskip \newline
\indent
 There are limits to the amount of energy which can extracted in this way. A particle crossing $H^+$ must have $ P_{\mu} (k^{\mu} + \Omega_H m^{\mu}) \leq 0$ as both are future directed causal curves. Defining $L = m\cdot P$, one has $E-\Omega_H L \geq0$. So the particle carries energy $E$ and angular mom $L$ into the BH. Hence $\delta M = E$ and $\delta J = L$. Our inequality gives $$\delta J \leq \frac{\delta M}{\Omega_H} = \frac{2M(M^2 + \sqrt{M^4 - J^2})}{J}\delta M$$ Defining $M_{irr} = \bigg(1/2 [M^2 +\sqrt{M^4 - J^2}]\bigg)^{1/2}$ we clearly see that $$M^2\geq M_{irr}$$ so there is a bound to how much energy can be extracted from the BH. One can show that $A = 16\pi^2 M_{irr}$.

\subsection{Anti de-Sitter Space}
Because one of our interests is superradiance in Kerr-AdS we give here a quick introduction into Anti de-Sitter space and its properties.\bigskip
\newline \indent 
The simplest vacuum solution of Einstein's equation with cosmological constant,
$$G_{ab} + \Lambda g_{ab} = 0$$ are spacetimes of constant curvature. They are locally characterized by the condition $$R_{abcd} =  \frac{R}{(d-1)d}(g_{ac}g_{bd}-g_{ad}g_{bc})$$ where d is the dimension of spacetime. Making use of these expressions we see that$$G_{ab} = -Rg_{ab} \frac{d-2}{2d} = -\Lambda g_{ab} $$ hence the Ricci scalar is the constant $R = \frac{2d}{d-2}\Lambda$. We are in the domain of AdS when $\Lambda<0$. \bigskip 
\newline \indent 
In $d=4$, which is our primary dimension on interest, the metric can be written as
 \begin{equation} \label{eq:adsmet}
ds^2 = -\bigg(1+\frac{r^2}{L^2}\bigg)dt^2 + \bigg(1+\frac{r^2}{L^2}\bigg)^{-1}dr^2 + r^2(d\theta^2 + \sin^2\theta d\phi^2)
\end{equation}
 The quantity $L$ in eq \ref{eq:adsmet} is the radius of curvature of the spacetime and is related to $\Lambda$ by $$\Lambda = -3/L^2 .$$ This is a maximally symmetric solution to Einstein's equation. \bigskip
\newline \indent
One of the most famous properties of AdS, and one which is central to the discussion of the superradiant instability later on, is that it acts like a ``box". To demonstrate this consider the norm of the tangent vector of a radially outward null geodesic: $$0 = ||\dot{x}||_{\text{AdS}} = g_{\mu\nu}\dot{x}^\mu\dot{x}^\nu= -\bigg(1+\frac{r^2}{L^2}\bigg)\dot{t}^2 + \bigg(1+\frac{r^2}{L^2}\bigg)^{-1}\dot{r}^2$$ from this we have $$\frac{dt}{dr}= \frac{1}{1+r^2/L^2} \implies \Delta t=\int^\infty_0 \frac{dr}{1+\frac{r^2}{L^2}} = \frac{\pi}{2}L$$
We see that it takes a finite time for a radial null geodesic to reach the boundary and so we can think of AdS in some sense as an enclosed space. Note that it follows from this that AdS  is not globally hyperbolic. For any hypersurface one can always construct a timelike curve which reaches the boundary before it is able to intersect the surface. Hence when evolving AdS initial data, boundary conditions (BCs) become very important.

\subsection{Perturbations of Kerr Black Holes}
Black Hole perturbation theory is an incredibly complex and rich subject. Here we will simply introduce what is relevant to the Kerr BH. For convenience we have left out a satisfactory discussion of Newman-Penrose (NP) formalism. For those unfamiliar, we highly recommend that the reader consult chapters 2,6 and 7 of \cite{chandra}.

Newman-Penrose formalism is a tetrad formalism in which the basis vectors are selected so as to emphasize the lightcone structure of the space-time. We pick an "isotropic tetrad" $$\{e_1,e_2,e_3,e_4\}=\{l,n,m,\bar{m}\}$$ with $l,~n$ real valued and $m$ complex valued, such that the only non-zero inner products are $$l^{\mu} n_\mu =1=-m^\mu \bar{m}_\mu$$

Given a tensor $T_{ij}$ we can project onto the tetrad frame and express the object in tetrad coordinates: $$T_{ab} = e_a^ie_b^jT_{ij}~~.$$ We can pass freely in either direction, considering the problem in which ever frame provides the most simplification. The original metric of the spacetime can be recovered via $$g_{\mu\nu} = 2[l_{(\mu}n_{\nu)} - m_{(\mu}m_{\nu)}]$$

For electromagnetic and gravitational perturbations we can consider the relevant tensors $F_{\mu\nu}$ and $C_{\alpha\beta\gamma\delta}$ in the tetrad frame. Because $m, \bar{m}$ are complex valued we can actually express the $6$ and $10$ independent components of the above tensors as a set of $3$ and $5$ complex valued scalars respectively. The information contained in the Maxwell tensor is encoded in the following 
\begin{equation}
\varphi_0 = F_{\mu\nu}l^{\mu}m^{\nu},~~ \varphi_1 = \frac{1}{2}F_{\mu\nu}(l^{\mu}n^{\nu} - m^{\mu}\bar{m}^{\nu}),~~ \varphi_2 = F_{\mu\nu}\bar{m}^{\mu}n^{\nu}
\end{equation}
and the Weyl tensor is distilled into the 5 complex scalars 
\begin{eqnarray*}
\Psi_0& = &-C_{1313} = C_{\alpha\beta\gamma\delta} l^\alpha m^\beta l^\gamma m^\delta\\
\Psi_1& = &-C_{1213} \\
\Psi_2& = &-C_{1342} \\
\Psi_3& = &-C_{1242} \\
\Psi_4& = &-C_{2424} = - C_{\alpha\beta\gamma\delta}n^\alpha \bar{m}^\beta n^\gamma \bar{m}^\delta
\end{eqnarray*}

Maxwell's equations manifest in the NP formalism as a system of equations involving $\varphi_i$, the derivative operators given by the tetrad basis, $e_i = e_i^\mu \partial_\mu$, and a set of 12 \textit{spin coefficients} which are related to the structure constants of the tetrad basis under the bracket operation.
The Weyl tensor shares the symmetries of the Riemann tensor and has the further restriction of tracelessness.This gives a similar set of equations this time involving $\Psi_i$ in place of $\varphi_i$. \bigskip \newline \indent
The form of these tetrads, corresponding to the Kerr geometry, was discovered by Kinnersly and is given by: 
\begin{equation}\label{eq:kin}
l = \bigg(\frac{r^2+a^2}{\Delta}, 1, 0, \frac{a}{\Delta}\bigg)
\end{equation}
$$n=\frac{1}{2(r^2 +a^2\cos^2\theta)} \bigg(r^2+a^2 -\Delta,0,a\bigg) $$
$$m = \frac{1}{\sqrt{2}(r+ia\cos\theta)}\bigg(ia\sin\theta,0,1,\frac{i}{\sin\theta}\bigg) .$$
\bigskip \newline \indent
In the effort to obtain linearized perturbation equations a natural first approach would be to start with the Einstein equation and let $g_{\mu\nu} \to g_{\mu\nu} + h_{\mu\nu}$ for metric perturbation $h_{\mu\nu}$. Expanding the field equations to first order in $h_{\mu\nu}$ yields a set of linear equations. In the setting of the Kerr geometry however this approach is complicated. The fewer symmetries, relative to say the Schwarzchild solution, mean that the resulting PDEs in $r$ and $\theta$ are not seperable.

Fortunately the NP formalism provides a simpler alternative approach.
It can be shown that when studying electromagnetic \cite{ipser} and gravitational \cite{wald} perturbations of the Kerr geometry it suffices to consider the NP scalars $\{\varphi_0,\varphi_2\}$ and $\{\Psi_0,\Psi_4\}$ respectively. Further, it was shown by Teukolsky \cite{Teuk} that the linear perturbations of the Kerr BH could be described by a single master equation: 
\begin{equation}\label{eq:mast}
\bigg[\frac{(r^2+a^2)^2}{\Delta}-a^2\sin^2\theta\bigg] \partial_t^2\psi + \frac{4Mar}{\Delta}\partial_t\partial_\phi \psi +\bigg[\frac{a^2}{\Delta}-\frac{1}{\sin^2\theta}\bigg]\partial_\phi^2 \psi
\end{equation} 
$$ -\Delta^{-s}\partial_r\bigg[\Delta^{s+1}\partial_r\psi\bigg] - \frac{1}{\sin\theta}\partial_\theta\bigg[\sin \theta \partial_\theta \psi\bigg]-2s\bigg[\frac{a(r-M)}{\Delta}+\frac{i\cos\theta}{\sin^2 \theta}\bigg]\partial_\phi \psi$$
$$-2s\bigg[\frac{M(r^2-a^2)}{\Delta} -r-ia\cos\theta\bigg]\partial_t \psi +(s^2\cot^2\theta - s)\psi=0 $$
with $\psi$ and $s$ are related as follows:
\begin{table}[h!]
	\begin{tabular}{ |c|c|c|c| } 
		\hline
		   $s$ & $0$ & $(1,-1)$ &$(2,-2)$ \\ 
		$\psi$ & $\Phi$ & $(\varphi_0, \rho^{-2}\varphi_2)$ & $(\psi_0,\rho^{-4}\psi_4)$ \\ 
		 	\hline
	\end{tabular}
	\centering
	\end{table}
	
where $\rho = -1/(r-ia\cos\theta)$. \bigskip \newline \indent
Further, by Fourier decomposing $\psi$ with the form $$\psi = \frac{1}{2\pi}\int d\omega e^{-i\omega t}e^{i\omega \phi} S(\theta)R(r)$$ Teukolsky was able separate eq \ref{eq:mast} into the following ODEs for $R$ and $S$:

\begin{equation}\label{eq:rode}
\Delta^{-s}\frac{d}{dr}\bigg(\Delta^{s+1}\frac{dR}{dr}\bigg) + \{[(r^2+a^2)^2\omega^2-4aMm\omega r +a^2m^2 +2ia(r-M)ms
\end{equation}
$$-2iM(r^2-a^2)\omega s]\Delta^{-s} + 2i\omega r s -\lambda\}R = 0 $$
\begin{equation}\label{eq:sode}
\frac{1}{\sin\theta} \frac{d}{d\theta}\bigg(\sin\theta \frac{dS}{d\theta}\bigg) -\bigg(a^2\omega^2\sin^2\theta + \frac{m^2}{\sin^2\theta} +2a\omega s\cos\theta + 
\frac{2ms\cos\theta}{\sin^2\theta}
\end{equation}
$$ + ~s^2\cot^2\theta -s\bigg)S + \lambda S=0$$ 

The separation constant $\lambda$ is constrained when BCs are imposed leading to a complex eigenvalue problem. 

\subsection{Superradiance}

We will now outline the theory of superradiant scattering of test fields on a BH background. For concreteness and simplicity we will consider an asymptotically flat spacetime (so not AdS). It should be noted that fluctuations of order $\mathcal{O(\epsilon)}$ in the scalar fields induce a change of order $\mathcal{O}(\epsilon^2)$, so one is justified in fixing the BH geometry.

Let us assume that our spacetime is stationary and axisymmetric, as in the case of the Kerr BH. As we have seen above for such a spacetime various types of perturbations can be expressed in terms of a master variable $\Psi$. It can be shown that $\Psi$ obeys a Schrodinger-like equation 
\begin{equation} \label{eq:schro}
\frac{d^2 \Psi}{dr_\ast ^2} + V_{eff} \Psi = 0
\end{equation}
 with $V_{eff}(r)$ dependent on th curvature of the background and the test field properties. We let $r_\ast$ be some coordinate which maps $[r_+,\infty] \to \mathbb{R}$. Consider the scattering of a monochromatic wave of frequency $\omega$ with $t ~~\&~~ \phi$ dependence given (because of the $\partial_t ~~\&~~ \partial_\phi$ isometries) by $e^{-i(\omega t - m\phi)}$. Supposing $V_{eff}$ is constant at the boundaries, the asymptotics of eq \ref{eq:schro} give
\begin{equation} \label{eq:supas}
\centering
\Psi \sim 
\begin{cases}
\mathcal{T}e^{-ik_+ r_\ast} & \text{as}~~ r \to r_+ \\
\mathcal{R}e^{ik_\infty r_\ast} + \mathcal{I}e^{-ik_\infty r_\ast} &\text{as} ~~r\to \infty 
\end{cases}
$$	
$$\end{equation}

where $k^2_+ = V_{eff}(r\to r_+)$ and $k^2_\infty = V_{eff} (r\to \infty)$. The event horizon imposes the boundary condition of a one-way membrane. We have a wave incident from spatial infinity of amplitude $\mathcal{I}$ which upon reaching the boundary at $r_+$ gives rise to a transmitted wave of amplitude $\mathcal{T}$ and reflected wave of amplitude $\mathcal{R}$. Superradiance corresponds to the condition that $|\mathcal{R}|^2 > |\mathcal{I}|^2$

As a further simplification lets assume that $V_{eff}$ is real. The symmetries of the field equations imply that there is another solution $\bar{\Psi}$ satisfying the complex conjugate of the above BCs. Note $\Psi$ and $\bar{\Psi}$ are linearly independent which implies that their Wronskian, $W$, does not depend on $r_\ast$. Hence $-2ik_+|\mathcal{T}|^2 = W(r_+) = W(\infty) = 2ik_\infty(|\mathcal{R}|^2 - |\mathcal{I}|^2)$ which gives
$$|\mathcal{R}|^2 = |\mathcal{I}|^2 - \frac{k_+}{k_\infty}|\mathcal{T}|^2.$$ We see that superradiance occurs when $\frac{k_+}{k_\infty}<0$.

\section{Superradiance and the Kerr BH}
In this section we discuss superradiant scattering in the Minkowski background. As we have seen superradiant scattering, like the Penrose Process, is a means of extracting energy from a Kerr BH. For an incident wave of suitable conditions, reflection off of the event horizon occurs and the outgoing wave is more energetic (i.e has greater amplitude) than the ingoing wave. It should be emphasized that there is no change in frequency involved in superradaince; it is not a Doppler phenomenon. Waves of arbitrary spin may be considered by introducing the appropriate field term in the Einstein-Hilbert field action. For our purposes, as we are mainly reviewing the work of \cite{russ}, we will discuss waves of spin $s=0,~1 ~~\&~~2$, i.e. scalar, electromagnetic and gravitational perturbations. We will discuss some of the conditions for superradiance to occur. We will also study the magnitude of reflection and its dependence on the relevant quantities associated with the scattered wave.

\subsection{Perturbations of the Kerr Metric}
Consider for a moment and electromagnetic perturbation.
The total energy flux per steradian at infinity is given by $$\frac{d^2E}{dtd\Omega} = \lim_{r\to \infty} r^2T^r_t~.$$ Now the Maxwell tensor can be expressed in terms of the NP scalars as follows $$4\pi T_{ij} = \varphi_0\bar{\varphi_0}n_in_j + \varphi_2\bar{\varphi_2}l_il_j+2\varphi_1\bar{\varphi_1}[l_{(i}n_{j)}+m_{(i}\bar{m}_{j)}]$$ $$-4\bar{\varphi_0}\varphi_1n_{(i}m_{j)} -4\bar{\varphi_1}\varphi_2l_{(i}m_{j)} + \varphi_2\bar{\varphi_0}m_im_j  $$ $$+ ~~ \text{complex conjugate of the preceeding terms} .$$ Using the Kinnersly tetrad \ref{eq:kin} we have $$ T^r_t = -\frac{1}{4}\varphi_0 \bar{\varphi_0} + \varphi_2 \bar{\varphi_2}$$ with the first term corresponding to an ingoing wave and the second to an outgoing wave.
Thus we interpret the terms as 
\begin{equation}\label{eq:efux}
\bigg(\frac{d^2E}{dtd\Omega}\bigg)_{in} = \lim_{r\to \infty}\frac{r^2}{8\pi}|\varphi_0|^2, ~~\bigg(\frac{d^2E}{dtd\Omega}\bigg)_{out} = \lim_{r\to \infty}\frac{r^2}{2\pi}|\varphi_2|^2 .
\end{equation}
In the case of gravitational perturbations we can get at the desired energy fluxes by a similar method only with the use of the Landau-Lifshitz pseudotensor. The results obtained are 
\begin{equation}\label{eq:gflux}
\bigg(\frac{d^2E}{dtd\Omega}\bigg)_{in} = \lim_{r\to \infty}\frac{r^2}{64\pi\omega^2}|\Psi_0|^2,~~ \bigg(\frac{d^2E}{dtd\Omega}\bigg)_{out} = \lim_{r\to \infty}\frac{r^2}{4\pi\omega^2}|\varphi_4|^2 
\end{equation}

We need only consider the $s=1,~2$ scalars $\varphi_0$ and $\Psi_0$ as it turns out that  the $s=-1,~2$ follow from these. Recalling the discussion of the Teukolsky master equation we use the ansatz $$\varphi_0,~ \psi_0 = R(r)S(\theta)e^{im\varphi - i \omega t}$$ with frequency $\omega$ and angular momentum z-component $m$.

The $S(\theta)$ ODE \ref{eq:sode} combined with the physically desirable BCs, $|S(0)|<\infty$ and $|S(\pi)|<\infty$ yields an eigenvalue problem for $$\lambda=~~_{s}\lambda_l^m(a\omega)=~~_{s}\lambda_l^{-m}$$
 where $l$ is some whole number such that $l\geq \max(|m|,s)$. For $a\omega = 0$ (either Schwarzchild or wave of zero frequency) one has $_{s}\lambda_l^m(0) = (l -s)(l+s+1)$ and the eigenfunctions are weighted spinor spherical harmonics. For $a\omega \neq 0, ~~\lambda$ is not analytically expressible as a function of $l, n ~~\&~~ a\omega$.
 \newline \indent
 It remains to treat the $R(r)$ ODE \ref{eq:rode}. If we introduce the coordinate $y$ defined by $$\frac{dy}{dr}=\frac{r^2+a^2}{\Delta},~y\in(-\infty, \infty)$$ the asymptotic solutions of eq \ref{eq:rode} are given by
  \begin{equation}\label{eq:rbig}
  R(r\to\infty) \sim \mathcal{I}\frac{e^{-i\omega r}}{r} + \mathcal{R}\frac{e^{i\omega r}}{r^{2s+1}} 
  \end{equation}
  and 
  \begin{equation}\label{eq:rhor}
  R(r\to r_+) \sim\ \mathcal{T}\frac{e^{-i(\omega-m\Omega)y}}{\Delta^s}
  \end{equation}
  where $\Omega=\frac{a}{r_+^2 +a^2}$. Note that as $r\to \infty$ 
  \begin{equation}\label{eq:eqd}
   \varphi_0 \to \bigg(\mathcal{I}\frac{e^{-i\omega r}}{r} + \mathcal{R}\frac{e^{i\omega r}}{r^{2s+1}}\bigg)S_{in}(\theta)e^{im\phi - i\omega t}
   \end{equation}

\subsection{Conditions for Superradiant Scattering}
 We  now present the frequency conditions necessary for superradiance arrived at in \cite{russ}. We also discuss the dependence of the strength of the reflection $R$ on $\omega$ and the spin number $s$. \bigskip
\newline \indent 
Comparing equations \ref{eq:rbig} and \ref{eq:rhor} to the results of our discussion of superradiance in the introduction we see that $k_H = \omega -m\Omega$. The frequency dependent condition for superradiant scattering of an $m$-mode wave is thus
\begin{equation}\label{eq:supcond}
k_H =\omega - m\Omega<0
\end{equation}
 It follows that the condition is the same for all integral values of s and hence the same for scalar, vector and gravitational waves. The convention that $w>0$ and the observation that Schwarzchild solution corresponds to $\Omega = 0$ implies superradiant scattering doesn't occur for Schwarzchild BH. \newline
\newline \indent
As a check of physical plausibility it is a good idea to make sure the condition just stated adheres to the laws of BH thermodynamics. In particular, when superradiance occurs energy is extracted from the BH causing $M$ and $a$ to decrease. At first glance this might seem ominous as the surface area of the horizon grows monotonically with respect to $a^2$ but the $2_{nd}$ law of BH mechanics requires that $S_{hor}$ increase with time. We will show that $M$ and $a$ decrease in such a way that $S_{hor}$ actually increases under the process of superradiant scattering. \bigskip
\newline \indent
 Let $I$ be the energy flux of an incident wave of frequency $\omega$ and multipole order m. Then energy flux of the reflected wave is $RI$ and 
 \begin{equation} \label{eq:eqh}
 \frac{dM}{dt} = -(R-1)I, ~~ \frac{dL}{dt} = -\frac{m}{\omega}(R-1)I
 \end{equation}
  These expressions makes sense; $RI - I$ is just $\dot{E_f} - \dot{E_i}$ the energy flux gained by the wave which, is the negative of the energy flux lost by the BH (i.e $-\frac{dM}{dt}$). \bigskip
  \newline \indent
  Recall the discussion of the irreducible mass $M_{irr}$ in section 1; specifically the relation $$S_{hor} = 4\pi(r_+^2+a^2)=16\pi M_{irr}^2~.$$ It also follows from the definition of $M_{irr}$ that $$M^2 = M_{irr}^2 + L^2/{4M_{irr}^2} .$$ By considering the total derivative of $M(M_{irr}^2,L)$, making use of $\partial M/{\partial L} = L/{(M*4M_{irr}^2)}= aM/{M(r_+^2 + a^2)}= \Omega$ and eq \ref{eq:eqh}, it follows that $$\frac{dS_{hor}}{dt} = 16\pi \frac{d}{dt}M_{irr}^2 = 32\pi\frac{M}{1-a^2/r_+^2}(\dot{M} - \Omega\dot{L})$$ 
  \begin{equation}\label{eq:eqi}
  = 32\pi \frac{MI}{1-a^2/r_+^2}(\frac{n\Omega}{\omega}-1)(R-1) \geq 0 
  \end{equation}
  We see that the process is reversible, that is $dS/dt = 0$, only when $a<M$(ensures $r_1 \in \mathbb{R}$ and not extremal) and $\omega = m\Omega$. We will show below that this implies $R=1~~(\&~~ \dot{M} = 0)$. So reversibility corresponds to a perfectly reflected wave. It is apparent that we may get as close to reversibility as we wish by choosing $\omega$ arbitrarily close to $m\Omega$. \bigskip
  \newline \indent
  We will now study the behavior of $R$ in the neighborhoods of $\omega =0$ and $\omega = m\Omega$. The amplification factor $R$ can be determined numerically by integrating equations \ref{eq:rode} and \ref{eq:sode}. If we restrict attention to the low frequency realm the problem has also been solved analytically \cite{russ}. In what follows we simply give the analytically obtained expressions for $R$ without any prior derivation. See Appendix B of \cite{supover} if curious.

  For a wave of quantum numbers $s,~l,~m$ we have reflection coefficient $_sR_{lm}$. It can be shown that $R$ is of the from 
  \begin{equation}\label{eq:refl}
  _sR_{lm} -1 = (_0R_{lm}-1)\bigg[\frac{(l-s)!(l+s)!}{(l!)^2}\bigg]^2
  \end{equation}
  	with the reflection for the scalar wave given by 
 	$$_0R_{lm}-1 = -8Mr_+\big(\omega-m\Omega\big)\omega^{2l+1}(r_+-r)^2l \bigg[\frac{(l!)^2}{(2l!)(2l+1)!}\bigg]^2\prod_{k=1}^{l}\bigg[1+\frac{M^2}{k^2}\bigg(\frac{\omega-m\Omega}{\pi r_+ T_H}\bigg)\bigg]$$
 	where $T_H = \frac{r_+ -r_-}{4\pi r^2_+}$ is the temperature of the BH. The above expressions are valid in the region $a \leq M,~ \omega M \ll 1$ and for any spin $s$. Furthermore the expression is physically valid even when the superradiant condition $\omega<m\Omega$ is not satisfied. In that setting eq \ref{eq:refl} describes the absorption cross section of a rotating BH. \bigskip
 	\newline \indent
 	 Note that in $_sR_{lm} -1>0$ when $\omega<m\Omega$, for any $s ~~\&~~ l$. For a given $s$ we see that for $l\ll s^2$, $_sR_{lm}\approx~~ _0R_{lm}$. Restricting or focus to those $s =1,~2$ physically relevant case we see $$_1R_{lm} -1 = (_0R_{lm}-1)\bigg[\frac{(l+1)}{l}\bigg]^2$$ and $$_2R_{lm} -1 = (_0R_{lm}-1)\bigg[\frac{((l+1)(l+2)}{(l-1)l}\bigg]^2$$
 	so at most the electromagnetic and gravitational waves are amplified a factor of $4$ and $36$ times more than the corresponding $\{l,m\}$ scalar wave, respectively. Letting $\omega\to 0$, we only need keep the lowest order terms in $\omega$. Hence we see that for $m\ne 0,~~_sR_{lm} -1 \sim \omega^{2l+1}$.   \bigskip
 \newline \indent
  Now the $\omega \to  m\Omega$ case. Consider the quantity $\alpha = 1 - \frac{\omega}{m\Omega}$. One can show that if $a<M$ then $_sR_{lm} -1 \sim \alpha$ in the region $|\alpha|Q_1 = |\alpha|a\frac{m}{(r_1-r_2)} >0$. Further constraining $a\ll M ~~\&~~ n \ll \frac{M}{a}$ places us the realm of small $\omega$ allowing us to use eq \ref{eq:refl} compute the coefficient of $\alpha$.
  
  For the extremal Kerr BH, $a=M$, as $\omega \to m\Omega$ we have two cases. Let $$\delta^2 = 2m^2 - \lambda-(s+1/2)^2.$$ If $\delta^2<0$ then 
   \begin{equation} \label{eq:nast}
  _sR_{lm} -1 = 4\text{sgn}(\alpha) |\delta|^2(2m^2|\alpha|)^{2|\alpha|}\frac{|\Gamma(1/2 + s + |\delta|+im)|^2|\Gamma(1/2 - s + |\delta|+im)|^2}{\Gamma(1+2|\delta|)^4}e^{\pi m[1-\text{sgn}(\alpha)]}
  \end{equation}
  which is continuous and varies monotonically in the vicinity of $\omega = 0$. \newline \indent
  For $\delta^2>0$ on the other hand, in the region $|\alpha| \ll n^{-4} \max(|\alpha|^2, 1)$ we have: 
  \begin{equation} \label{eq:nast2}
  (_sR_{ln} -1)^{-1} = \frac{\text{sgn}(\alpha)e^{-\pi m[1-\text{sgn}(\alpha)]}}{\sinh(2\pi\delta)^2} \bigg\{\cosh(\pi[m-\delta])^2e^{-\pi \delta[1-\text{sgn}(\alpha)]} + \cosh(\pi[m+\delta])^2e^{\pi \delta[1-\text{sgn}(\alpha)]}
  \end{equation}
  $$-2\cosh(\pi[m-\delta])\cosh(\pi[m+\delta])\cos\big[\gamma_0 -2\delta\log(2m^2|\alpha|)\big] \bigg\}$$ 
  	where $\gamma_0$ is a function involving the argument of the $\Gamma$ terms. See \cite{russ} for the exact form. \bigskip \newline \indent
  Note that $\delta^2>0$ is  satisfied by the majority of modes. For example if $s=1$ it holds for all $l=m\geq1$ and if $s=2$ it holds for all $l=m=2$.
  In the vicinity of the onset of superradiant scattering, $\alpha =0$ the reflection coeffienct $R$ has an infinite number of oscillations in the region $|\alpha|m^2\ll1$. Aside from the case when $m=1 ~~\&~~ \pi\delta \leq 1$ these oscillations have small amplitude and can be ignored. In case $\alpha>0$ we have $$_sR_{ln} -1 \approx e^{2\pi(\delta - m)}$$ and the amplification factor is discontinuous near the onset of superradiance. For $\alpha<0,$ $\min~ _sR_{ln} =0$ suggesting that the barrier can be totally transparent, as one would expect for the region unable to superradiantly scatter. \bigskip
  \newline \indent
  Switching our attentions to the non-extremal Kerr BH, if $a\ne M$ but $M-a \ll M$ and $m\ll\sqrt{M/(M-a)}$ then $R(\delta,m,s,\alpha)$ is described by equations \ref{eq:nast} or \ref{eq:nast2}, depending on the values of sign of $\delta^2$, in the region $Q_1^{-1} << |\alpha| \ll m^{-1}$. In region $|\alpha|<<Q_1^{-1}, ~~ R-1 \sim \alpha$. Hence $R$ is continuous at $\alpha =0$ when $a\neq M$. 
  \newline \indent
  Using our expressions for $R$, calculations of the magnitude of $R$ yield: $R_{em}-1<.1$, in particular for $ a=M, \omega = \Omega-0$ we get $_1R_1^1 - 1 \approx .02$. For gravitational waves with $a=M, \omega=2\Omega-0$ we have $_2R_2^2 -1 =1.37$ so reflected gravitational waves can more than double in amplitude! \bigskip \newline \indent
  In general for fixed $s ~~\&~~ m \to\infty$ the effect decreases as an mth-power exponential; when $a=M,\omega=n\Omega -0 ~~\&~~ m\gg s^2$ we get $$_sR_{mm}-1 \approx e^{-m\pi(2-\sqrt{3})}$$

\section{Kerr-AdS and the Superradiant Instability}
Now we shift our focus by studying gravitational perturbations in a Kerr-AdS background. Because of the box like nature of AdS we see that superradiance will tend to lead to instabilities; a superradiantly reflected wave is free to bounce off the boundary and return to the ergoregion in finite time. We will see \cite{Santos} that general instabilities in Kerr-AdS are always superradiant in nature. Finally we will explore the possible evolution of the superradiantly perturbed Kerr AdS BH.
\subsection{Kerr-AdS}
In this section we give a brief overview of the properties of Kerr-AdS BHs.
For purposes of studying instability it is useful to use a variation of Boyer-Lundquist coordinates,$\{T,r,\theta,\varphi\}$, introduced by Chambers and Moss \cite{cham} given by $$\{t=\Xi~T,r,\chi = a \cos\theta, \phi \}$$ where $a$ is the rotation parameter of the solution, $L$ is the radius of curvature of the AdS background and $\Xi = 1-\frac{a^2}{L^2}$. In this coordinate system the Kerr-AdS metric is given by 
\begin{equation}\label{eq:AdSmet}
 ds^2 = -\frac{\Delta_r}{(r^2+\chi^2)\Xi^2}\bigg(dt-\frac{a^2-\chi^2}{a}d\phi\bigg)^2 + \frac{\Delta_\chi}{(r^2+\chi^2)\Xi^2}\bigg(dt-\frac{a^2+r^2}{a}d\phi\bigg)^2 
 \end{equation}
 $$+ \frac{(r^2+\chi^2)}{\Delta_r}dr^2 + \frac{(r^2+\chi^2)}{\Delta_\chi}d\chi^2$$ 
 where 
 $$\Delta_r = (r^2+a^2)\bigg(1 + \frac{r^2}{L^2}\bigg) -2Mr,~~ \Delta_\chi= (a^2-\chi^2)\bigg(1- \frac{\chi^2}{L^2}\bigg)~.$$ In this frame the horizon angular velocity and temperature are $$\Omega_H = \frac{a}{a^2+r_+^2},~~ T_H = \frac{1}{\Xi}\bigg[\frac{r_+}{2\pi}\bigg(1+\frac{r_+^2}{L^2}\bigg)\frac{1}{r^2_+ +a^2}-\frac{1}{4\pi r_+}\bigg(1-\frac{r_+^2}{L^2}\bigg)\bigg].$$ The Kerr-AdS BH asymptotically approaches global AdS with radius of curvature $L$. This is not obvious when one looks at eq \ref{eq:AdSmet} because the coordinate frame $\{t,r,\chi,\phi\}$ rotates at infinity with $\Omega_\infty = -a/(L^2 \Xi)$. If one introduces the coordinate change $$T=t/\Xi, ~~ \Phi = \phi + \frac{a}{L^2}\frac{t}{\Xi}$$ $$R = \frac{\sqrt{L^2(a^2+r^2)-(L^2+r^2)\chi^2}}{L^2\sqrt{\Xi}}, ~~\cos\Theta = \frac{Lr\sqrt{\Xi}\chi}{a\sqrt{L^2(a^2+r^2)-(L^2+r^2)\chi^2}}$$ and then considers the limit as $r\to \infty$ one gets $$ds^2 = -\bigg(1+ \frac{R^2}{L^2}\bigg)dT^2 + \frac{dR^2}{\bigg(1+ \frac{R^2}{L^2}\bigg)} + R^2(d\Theta^2 + \sin^2(\Theta)d\Phi^2) = ds^2_{\text{AdS}}$$ which we recognize from the section introducing AdS. Hence the conformal boundary of the bulk spacetime is the Einstein static universe $\mathbb{R}\times S^2: \lim_{R\to\infty} \frac{L^2}{R^2}ds^2_{\text{AdS}} = -dT^2 + d\Theta^2 + \sin^2\Theta d\Phi^2$. \bigskip
\newline \indent
The ADM mass and angular momentum of the BH are related to the parameters $M$ and $a$ by $M_{ADM} = M/\Xi^2 ~~ \& ~~ J_{ADM} = Ma/\Xi^2$. We can express the angular velocity and temperature in the manifestly globally AdS coordinates in terms of those obtained in Chambers-Moss (CM) coordinates: $$T_h = \Xi T_H ~~\&~~ \Omega_h = \Xi \Omega_H + a/L^2$$ As in the Kerr BH the event horizon is located at the largest real root of $\Delta_r,~~r=r_+,$ and is a Killing horizon generated by the KVF $K=\partial_T + \Omega_h \partial_\Phi$. We can express the mass parameter in terms of $a,~r_+$ and $L$ as follows $M =(r_+^2 +a^2)(r_+^2+L^2)/(2L^2r_+)$.\bigskip \newline \indent Any regular BH solution must obey $T_H\geq0 ~~\&~~ a/L<1$ which gives us restrictions on $r_+/L$ and $a/L$:
\begin{equation}
\frac{a}{L}\leq \frac{r_+}{L}\sqrt{\frac{L^2+3r_+^2}{L^2-r_+^2}}, ~~ \text{for} ~~ \frac{r_+}{L} < \frac{1}{\sqrt{3}}
\end{equation}
$$\frac{a}{L}<1,~~\text{for} ~~ \frac{r_+}{L} \geq \frac{1}{\sqrt{3}}$$
 In discussing superradiance it is useful to parametrize the BH by gauge invariant variables associated with its onset: $(R_+, \Omega_h)$, where $R_+ = \sqrt{r^2_+ + a^2}/\sqrt{\Xi}$. The extremal curve, where $T_H=0$, is given by $$|\Omega_h^{\text{extr}}| = \frac{1}{LR_+}\sqrt{\frac{(L^2+R^2_+)(L^2+3R_+^2)}{2L^2+3R^2_+}} .$$ Note $R_+$ is just the square root of the area of the spatial section of the EH divided by $4\pi$.

\subsection{Teukolsky Master Eq}
In general, the study of the linearized gravitational perturbations of the Kerr BH involves solving a coupled nonlinear PDE obtained from the linearized Einstein equation for the metric perturbation. This is hard to do. Fortunately, as we have already discussed, in $d=4$ the approach of Teukolsky simplifies the problem immensely. By studying gauge invariant scalar variables we can reduce the problem to solving a single PDE. Furthermore, by making use of harmonic decomposition when can make use of seperation of variables to further reduce the problem to two ODEs. \bigskip \newline \indent
 It should be noted that in the setting of AdS background the curvature slightly alters the terms in the ODEs \ref{eq:rode} and \ref{eq:sode}. Further, in asymptotically AdS we use the Chambers-Moss null tetrad 
 \begin{equation}
 l = \frac{1}{\sqrt{2}\sqrt{r^2+\chi^2}}\bigg(\Xi\frac{a^2+r^2}{\sqrt{\Delta_r}},\sqrt{\Delta_r},0,\frac{a\Xi}{\sqrt{\Delta_r}}\bigg),
 \end{equation}
 $$n =\frac{1}{\sqrt{2}\sqrt{r^2+\chi^2}}\bigg(\Xi\frac{a^2+r^2}{\sqrt{\Delta_r}},-\sqrt{\Delta_r},0,\frac{a\Xi}{\sqrt{\Delta_r}}\bigg) $$
 $$ m=\frac{-i}{\sqrt{2}\sqrt{r^2+\chi^2}}\bigg(\Xi\frac{a^2-\chi^2}{\sqrt{\Delta_r}},0, i\sqrt{\Delta_\chi},\frac{a\Xi}{\sqrt{\Delta_\chi}}\bigg)$$
  rather than the Kinnersly. Still, the information about gravitational perturbations with spin $s= -2$ is encoded in the perturbations of the Weyl scalar $\psi_4 = C_{abcd}n^a\bar{m_b}n^c\bar{m_d}$. The equation of motion for $\delta\psi_4$ is given by the Teukolsky master equation \cite{Teuk}. We expect something of the form:
$$\delta\psi_4 = (r-i\xi)^{-2} e^{-i\hat{\omega}t}e^{-im\phi}R^{(-2)}_{\hat{\omega}lm}(r)S^{(-2)}_{\hat{\omega}lm}(\chi) $$ where $S ~~\&~~ R$ satisfy 
\begin{equation} \label{eq:eqS}
\partial_\chi \bigg(\Delta_\chi\partial_\chi S^{(-2)}_{\hat{\omega}lm}\bigg) = -\bigg[-\frac{(K_\chi +\Delta^\prime_\chi)^2}{\Delta_\chi}+\bigg(\frac{6\chi^2}{L^2} + 4K^\prime_\chi + \Delta^{\prime \prime}_\chi\bigg) + \lambda\bigg] S^{(-2)}_{\hat{\omega}lm}
\end{equation}
and 
\begin{equation} \label{eq:eqR}
\partial_r \bigg(\Delta_r\partial_r R^{(-2)}_{\hat{\omega}lm}\bigg) =- \bigg[\frac{(K_r -i\Delta^\prime_r)^2}{\Delta_r}+\bigg(\frac{6r^2}{L^2} + 4iK^\prime_r + \Delta^{\prime \prime}_r\bigg) - \lambda\bigg]  R^{(-2)}_{\hat{\omega}lm}
\end{equation} 
where $$K_r = \Xi[ma - \hat{\omega}(a^2+r^2)], ~~K_\chi= \Xi[ma-\hat{\omega}(a^2-\chi^2)]$$
In the eigenfunctions $S_{\hat{\omega}lm}^{(-2)}(\chi)$ we have the spin weighted $s=-2$ AdS spheroidal harmonics. The positive integer $l$ specifies the number of zeros of $S$ along the polar direction, the value is $l - \max(|m|,|s|)$. Note that for the eigenfunction of interest to us,$~~S^{(-2)}$, $l_{\min} = |s| = 2$ and $-l\leq m \leq l.$ \bigskip 
\newline \indent
These equations implicitly contain 5 parameters $\{a,r_+,\hat{\omega}, m,l\}$. Considering a particular Kerr BH amounts the fixing $\{a,r_+\}$. To study the physical problem of interest we need to solve equations \ref{eq:eqS} and \ref{eq:eqR} but we also need to impose BCs to restrict the solutions to those which are physically meaningful. As a natural example; at infinity we want the perturbations to preserve global AdS.
\newline \indent
 At the horizon, it is not possible to have waves coming out from $r<r_+$, hence the BC is such that only ingoing modes are allowed. A Frobenius analysis at the horizon gives two independent solutions: $$R_{\hat{\omega}lm}^{(-2)} \sim A_{in}(r-r_+)^{1-i\frac{\bar{\omega}-m\Omega_H}{4\pi T_H}}[1+\mathcal{O}(r-r_+)] + A_{out}(r-r_+)^{-1+i\frac{\hat{\omega}-m\Omega_H}{4\pi T_H}}[1+\mathcal{O}(r-r_+)] .$$
One can extend the solution through the horizon by introducing ingoing Eddington-Finkelstein coordinates$\{v,r,\chi,\varphi\}$ defined by $$t=v - \Xi\int\frac{r^2+a^2}{\Delta_r}dr,~~ \phi = \varphi - \int\frac{a\Xi}{\Delta_r}dr~~.$$ Imposing the BC then amounts to requiring that the metric perturbation is regular in these ingoing EF coordinates. It follows \cite{BC} that this is the case iff $R(r)|_H$ behaves as $R(r)|_H \sim R_{IEF}(r)|_H(r-r_+)^{1-i\frac{\hat{\omega}-m\Omega_H}{4\pi T_H}}$ for a smooth function $R_{IEF}(r)|_H$. Thus the appropriate boundary condition yields $$R_{\hat{\omega}lm}^{(-2)} \sim A_{in}(r-r_+)^{1-i\bar{\omega}}[1+\mathcal{O}(r-r_+)] $$ where $\bar{\omega} = \frac{\hat{\omega}-m\Omega_H}{4\pi T_H}$, note the relevance of the sign of this quantity to superradiance.
\newline\indent
Shifting our attentions to the boundary at infinity, a Frobenius analysis of the radial Teukolsky yields $$R_{\hat{\omega}lm}^{(-2)}|_{r\to \infty} = B_+^{(-2)}L/r + B_-^{(-2)}L^2/r^2 + \mathcal{O}(L^3/r^3)~~.$$ We are interested in perturbations which preserve the asymptotic global AdS background. As shown in \cite{BC} the following Robin BC ensures this preservation: 
\begin{equation} \label{eq:robbc}
 B_-^{(-2)} = i\beta B_+^{(-2)}
 \end{equation}
  where $\beta$ has two possible values $\beta_s ~~\&~~ \beta_v$ for "scalar" and "vector" sector perturbations respectively. It should be emphasized that the terms scalar and vector do not refer to $s=0,~1$ perturbations. In the limit $a=0$ one can map the solutions of the Kerr-AdS perturbations as given by the Teukolsy formalism to the perturbations of the AdS-Schwarzschild background. When this is done it turns out the solutions with the $\beta_s$ BC correspond to the scalar harmonics and those with the $\beta_v$ the vector harmonics. \bigskip
\newline\indent 
As noted previously the CM coordinates, $\{t,r,\chi,\phi\}$, rotate at infinity. The coordinates $\{T,R,\Theta,\Phi\}$ are better suited to discussing the global AdS structure of the background at the boundary. Remember that the reason it is sufficient to solve equations \ref{eq:eqR} and \ref{eq:eqS} is because in CM frame $\partial_t ~~\&~~ \partial_\phi$ are isometries of the background geometry so any linear perturbation can be Fourier decomposed in these directions as $e^{-i\hat{\omega}t}e^{im\phi}$. \newline \indent
In the $\{T,R,\Theta,\Phi\}$ frame one measures frequency $\omega \equiv \Xi\omega + m\frac{a}{L^2}$ with perturbation decomposition $e^{-i\omega T}e^{im\Phi}$. The quantity $\omega$ can be viewed as the natural frequency as it measures the frequency wrt a frame which does not rotate at infinity. We will often refer to $\omega$ in plots because of its natural physical significance. In particular note that we can express $\bar{\omega}$ in terms of $\omega$ by way of $$\bar{\omega} = \frac{\Xi}{\Xi}\bigg(\frac{\bar{\omega+} + ma/(\Xi L^2) -m\Omega_H -ma/(\Xi L^2)}{4\pi T_H}\bigg) = \frac{\omega - m\Omega_h}{4\pi T_h}$$ where only quantities measured in $\{T,R,\Theta,\Phi\}$ appear in the above expression.

\subsection{QNMs and Superradiance in AdS}
Recall that for fixed $\{a,r_+\}$ (or equivalently $\{R_+/L,\Omega_hL\}$) and quantum numbers $\{l,m\}$, the equations \ref{eq:eqR} and \ref{eq:eqS} along with our BCs give us a complex eigenvalue problem for $\lambda$. The radial and angular ODEs are coupled through $\hat{\omega}$ and $\lambda$ and cannot be solved analytically when $M,~a\ne 0$. In \cite{Santos} numerical methods were used to find solutions of equations \ref{eq:eqR} and \ref{eq:eqS} subject to the scalar and vector BCs \ref{eq:robbc}. As with the Minkowski background, there is a region where an approximate analytical solution can also be obtained for the frequency spectrum. This analytical treatment is valid when we have a small horizon radius and still smaller rotation parameter $$ \frac{a}{L} \ll \frac{r_+}{L} \ll 1$$ and for perturbations with wavelength much bigger then the BH length scales or equivalently the low frequency limit (which we recognize from the treatment of the Minkowski background case). \bigskip
\newline \indent
 We now discuss some of the numerical results obtained in \cite{Santos}. Consider a Kerr-AdS BH lying somewhere in the phase diagram given by $\{R_+/L, \Omega_h L\}$. The stability of a generic perturbation $\delta \psi_4 \sim e^{-i\omega T}$ is clearly dictated by the sign of $\Im(\omega)$. In the case $\Im(\omega)<0$ we have a decaying perturbation, i.e a Quasinormal mode or QNM. Unstable modes on the other hand, grow exponentially and have $\Im(\omega)>0$. We also have the case in which the imaginary part vanishes, $\Im(\omega) = 0$. The amplitude of such a perturbation neither decays or grows with time. For a given pair $\{l,m\}$ in parameter space $\{R_+/L, \Omega_h L\}$ we can plot the onset curve (OC) of points for which the mode $\Im(\omega)=0$. Essentially for a given value of $R_+$ we find the value of $\Omega_h$ for which the eigenfrequency of equations \ref{eq:eqR} and \ref{eq:eqS} has $\Im(\omega)=0$. We trace out a curve in phase space of Kerr BHs which admit an $\{l,m\}$ mode with an amplitude constant in time. Note that this, a priori, tells us nothing about $\Re(\omega)$\bigskip

To better understand the nature of the unstable perturbations it is helpful to consider the real part of $\omega$ or more specifically $\Re(\omega) - m\Omega_h$. Recall that this quantity determines the sign of the energy flux through $\mathcal{H_+}$. In particular, superradiant modes have negative energy flux at the future horizon. Vanishing flux, perfect reflection in other words, occurs when $\Re(\omega)=m\Omega_h$. In \cite{Santos} the (numerically) obtained spectrum of $\omega$ revealed the following relationships between the real and imaginary parts of $\omega$. It was found that perfect reflection occurred whenever $\Im(\omega)=0$ and that $\Re(\omega)-m\Omega_h <0$ whenever $\Im(\omega)<0$. Hence unstable modes in Kerr-AdS are unstable precisely because of the superradiant instability. \bigskip

\begin{figure}[h!] 
	\includegraphics[width =16cm]{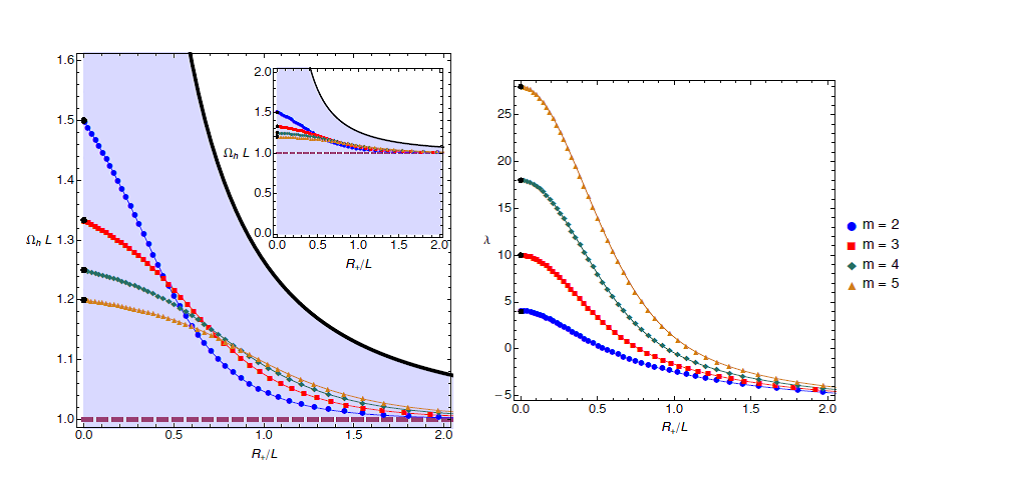}
	\caption{Phase diagram from \cite{Santos} of the Kerr-AdS BH. The curves are the OCs for the $l=m=2,~3,~4,~5$ scalar modes.}
	\centering
	\label{fig:fg2}
\end{figure}

In fig \ref{fig:fg2} we plot the onset curves of the first few $l=m$ "scalar" modes in the phase space of $\{R_+/L,\Omega_hL\}$. In the $\{R_+/L, \Omega_hL\}$ phase space each point in the blue shaded region, bounded above by the extremal curve (black), represents a Kerr BH which is stable when no perturbation is present. For a given $l=m$ scalar perturbation, all BHs above the $l=m$ onset curve are unstable to that particular type of perturbation. That is to say if a point lies above an $\{l,m\}$ onset curve then the $\{l,m\}$ mode eigenfrequency for that BH will be such that $\Im(\omega)>0$ and $\Re(\omega)<m\Omega$. On the other hand points beneath the $\{l,m\}$ onset curve are stable and such perturbations manifest as QNMs with $\Im(\omega)<0$ and $\Re(\omega) >m\Omega$. \bigskip
\newline \indent 
As $R_+/L\to 0$ we consider the BH as it becomes very small, approaching the global AdS limit in which the BH disappears. In the global AdS limit the scalar mode frequencies and angular eigenvalues can be computed analytically. Hence by following the OCs back to $R_+ = 0$ and comparing we get a good check of the numerics used.

Analytically the global AdS eigenfrequencies are given by:
\begin{equation} \label{eq:ws}
L\omega_s^{\text{AdS}} = 1 +l + 2p, ~~ \lambda = l(l+1)-2 .
\end{equation}
 where $p= 0,~1,~2,...$ is the number of radial nodes(the radial overtone). To obtain the values $\Omega_h|_{R_+=0}$ we use the superradiant onset condition to find $\Omega_h|_{R_+=0} = \omega_s^{\text{AdS}}/m$ and we set $p=0$ and $l=m$ giving $$L\Omega_h|_{R_+ = 0} = 1 + 1/m$$ As mentioned all the OCs plotted have no zero radial overtone, for each pair $\{l,m\}$ there is actually an OC for each value of $p$ but $p>0$ curves always lie above the $p=0$ curve. Hence the $p=0$ modes are the first to become unstable as the rotation $\Omega_h$ is increased. \bigskip

Note that all of the onset curves monotonically approach $L\Omega_h=1$ from above in the asymptotic limit $R_+/L \to \infty$, bunching up in process. Hence only BHs with $L\Omega_h>1$ can be superradiantly unstable as was previously convincingly argued by Hawking and Reall \cite{hawk}. A visually suggestive aspect of fig \ref{fig:fg2} should also be addressed. There is a value of $R_+$ where the $l=2=m$ curve dips below all the others and remains below as $R_+ \to \infty$. This might seem to suggest that there exists a (non compact) region in phase space which is superradiantly unstable to $l=2=m$ but stable to all other superradiant modes. In reaching this conclusion we have jumped the gun however. Each onset curve is monotone decreasing with respect to $R_+/L$ and the $L\Omega_h|_{R_+=0}$ decreases monotonically towards $1$ with respect to $m$, as we previously noted. Thus for any point $\{L\Omega_h^\prime,R_+^\prime/L\}$ on the $l=2=m$ onset curve we can find an $m$ such that for the curve $l=m, ~~ 1<L\Omega_h|_{R_+=0}<L\Omega_h^\prime$ which clearly implies the BH associated with $\{L\Omega_h^\prime,R_+^\prime/L\}$ is superradiantly unstable to perturbations $l=m$. \bigskip
\newline \indent
\begin{figure} [h!]
	\includegraphics[width = 14cm]{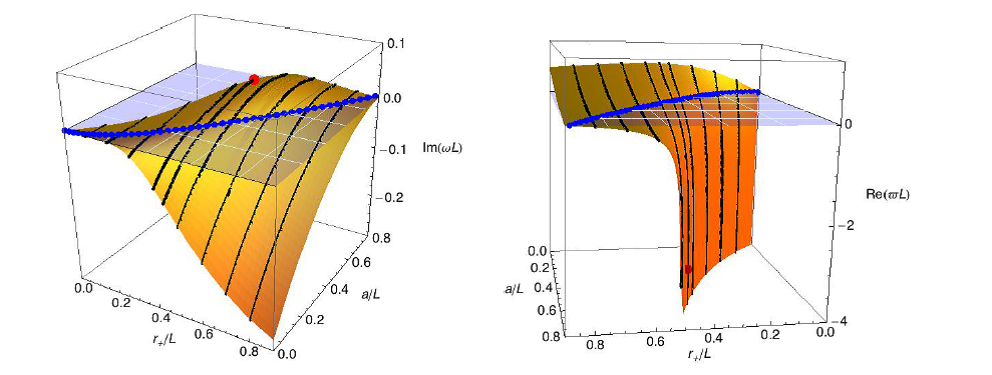}
	\centering
	\caption{A plot from \cite{Santos} of the $\Im(\omega L)$ and$ \frac{\Re(\omega)-m\Omega_hL}{4\pi T_h}$ with respect to $r_+/L$ and $a/L$ for the $l=2=m$ scalar mode. The point of maximal $\Im(\omega)$ and hence maximal growth rate is indicated by the red dot. It occurs at $\{r_+/L,a/L\}_{\max} = \{0.445\pm 0.020,~~0.589\pm 0.020\}$ where $\omega L \sim 1.397 + 0.032i$.}
	\label{fig:fg3}
\end{figure} \newline \indent
In order to further explore the stability properties of these BHs it is illuminating to consider a specific scalar mode. In \cite{Santos} the mode $l=2=m$ was considered. We will follow their lead and discuss the results they obtained. In fig \ref{fig:fg3} the $l=2=m$ eigenfrequency $\omega = \omega(r_+,a)$ is considered; $\Im(\omega L)$ and $\Re(\bar{\omega}L)$ are plotted with respect to $r_+/L$ and $a/L$. Recall that $Re(\bar{\omega})$ and not $Re(\omega)$ is what is relevant to superradiance. In the figure the 2-d surface formed by the plot has been marked to indicate physically important regions: an auxiliary plane marks the $\Re(\bar{\omega}) = 0 ~~\&~~ \Im(\omega)=0$. The black curves indicate paths of constant $r_+$. The onset curve is shown in blue. In the $\Im(\omega)$ plot the points on the surface above the auxiliary plane correspond to superradiantly unstable modes and in the $\Re(\bar{\omega})$ plot the superradiant modes are those below the plane. 
\newline \indent The red dot in fig \ref{fig:fg3} representing the BH most susceptible to $l=2=m$ mode perturbations corresponds to the point $\{R_+/L, \Omega_hL\} \sim \{0.914,1.295\}$ in fig \ref{fig:fg2}. Note that this occurs close to extremality but not at it. In fact at the onset of instability the timescale increases until achieving a maximum near extremality and then decreases as the $T_H = 0$ Kerr-AdS BH is approached.
\begin{figure}[h!]
	\includegraphics[width = 16cm]{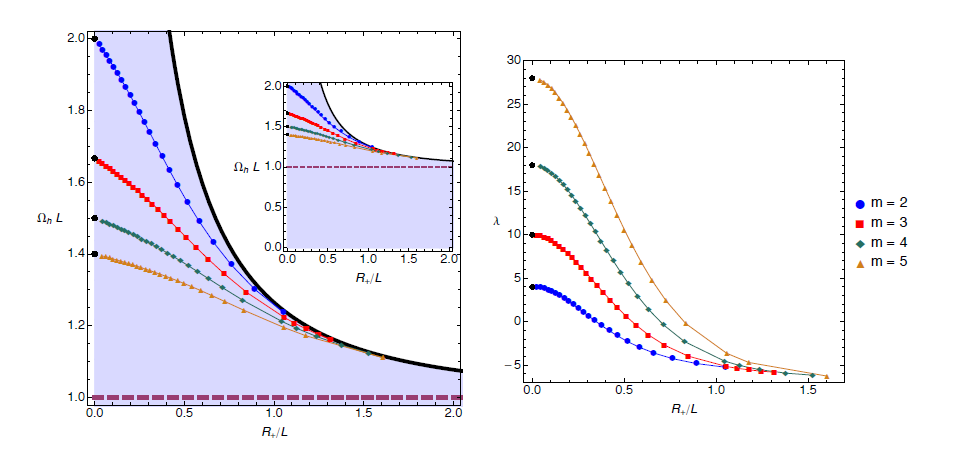}
	\centering
	\caption{Phase diagram of the Kerr-AdS BH from \cite{Santos}. The curves represent the OCs for the $l=m=2,~3,~4,~5$ vectors modes of a gravitational perturbation.}
	\label{fig:fg4}
\end{figure}
\bigskip \newline \indent
We can now consider gravitational vector modes obeying the BC with $\beta = \beta_v$. Consider fig \ref{fig:fg4}. As with the scalar modes the values of the onset curves at $R_+/L=0$ describe the vector normal modes of global AdS limit:
\begin{equation}\label{eq:wv}
L\omega^{\text{AdS}}_v = 2+l+2p,~~ \lambda = l(l+1)-2 .
\end{equation}

 In conjunction with the superradiant onset condition $\Omega_h|_{R_+=0} = \omega_v^{\text{AdS}}/m$, with $p=0 ~~ \&~~l=m$, we obtain $$L\Omega_h|_{R_+=0} =1+2/m .$$ As in the scalar case, the vector onset curves are bounded below with the condition $\Omega_hL>1$. Unlike the scalar case though, the vector OCs never cross one another and asymptotically approach the extremal curve. If a BH is unstable to the $l=2=m$ modes it is necessarily unstable to all $l=m\geq 3$ modes. The value of $R_+/L$ at which the OC hits extremality increases monotonically with $l=m$, the curves then both approach the line $\Omega L =1$. As $m\to \infty$ extremality is obtained only as $R_+/L \to \infty$. \bigskip
\begin{figure}[h!] \label{fig:fg5}
	\includegraphics[width = 14cm]{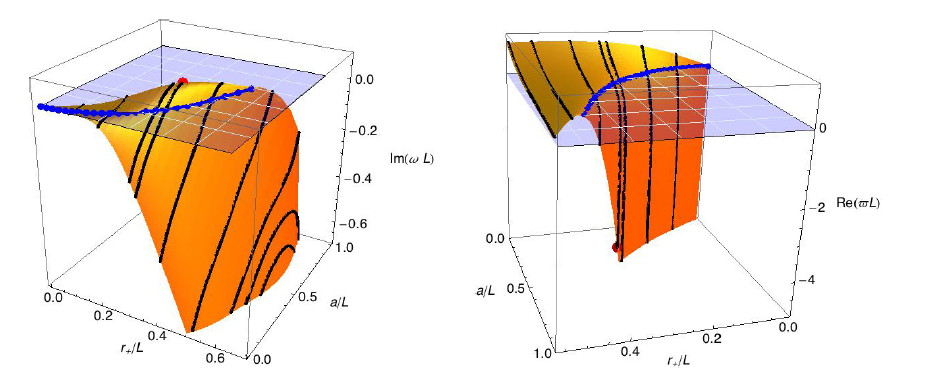}
	\centering
	\caption{Plot from \cite{Santos} of the $\Im(\omega L)$ and$ \frac{\Re(\omega)-m\Omega_hL}{4\pi T_h}$ wrt $r_+/L$ and $a/L$ for the $l=2=m$ vector mode. The point of maximal $\Im(\omega)$ and hence maximal growth rate is indicated by the red dot. It occurs at $\{r_+/L,a/L\}_{\max} = \{0.325\pm 0.020,~~0.386\pm 0.020\}$ where $\omega L \sim 2.667 + 0.058i$.}
	\label{fig:fg5}
\end{figure}
\newline\indent
As was done in the scalar case we consider the specific case $l=2=m$ for vector modes. Again an auxiliary plane divides the surface into the stable ($\Im(\omega)<0 ~~\&~~ \Re(\omega)>m\Omega$) and unstable ($\Im(\omega>0 ~~\&~~ \Re(\omega)<m\Omega$) BHs. The red dot in fig \ref{fig:fg5} representing the BH most susceptible to $l=2=m$ mode perturbations corresponds to the point $\{R_+/L, \Omega_hL\} \sim \{0.539,1.687\}$ in fig \ref{fig:fg4}. Note that moving along a constant $r_+/L$, the maximum of the vector superradiant instability is achieved much closer to the extremal curve than in the scalar case.
\newline\indent
The instability growth rate of scalar and vector modes is of the same order, with the vector rate being approximately twice as large as the scalar rate. Comparing the most unstable case in the vector and scalar modes we see that the BH corresponding to maximal vector instability is smaller (in terms of $R_+/L$) but rotates faster than corresponding scalar BH.

\subsection{Bifurcation at the onset of Superradiance}
We have seen that the Kerr-AdS BH is not stable to superradiance. For waves with the appropriate $\omega_R$ and $\omega_I$ conditions amplification occurs until $\Omega_h$ is sufficiently reduced by which time the mode has accumulated enough energy to backreact with the Kerr-AdS background. A natural question to ask then is what does this instability evolve into? 
\newline \indent
As we have already seen, by appealing to monotonicity, any Kerr-AdS BH with $\Omega_hL>1$ is superradiantly unstable to some gravitational perturbational mode $\{l,m\}$. For a given $m$ at the onset of superradiance we have an exact zero mode with $\Im(\omega) = 0$ and $\Re(\omega) = m\Omega_h$:  since the perturbation is proportional to $e^{-i\omega T+im\Phi}$ the amplitude of the zero mode is constant with respect to $T$.
These perturbational modes are of the form  $e^{-i\omega T+im\Phi}= e^{-im\Omega_hT + im\Phi} $  and so have the special property of being invariant under the horizon generating KVF, $K=\partial_T + \Omega_h\partial_\Phi$. \bigskip
\newline \indent
For a given $m$ it was suggested \cite{Reall} that the OC of the instability should initiate the merger of the Kerr-AdS BH with a new family of BH solutions, stable to $m$ superradiant modes and invariant under a single KVF, $K=\partial_T + \Omega_h \partial_\Phi$. This has been explored for the case of scalar field perturbations. BHs with a similar helical KVF that merge with the Kerr-AdS were found to have scalar hair orbiting the central core. For those unfamiliar, "hairy black hole" is a blanket term referring to a BH with a characterizing parameter other than $\{a,M,\mu\}$, i.e a BH which does not belong to the Kerr family. With the scalar field case as motivation we expect that an analogous family of single KVF BHs with "lumpy gravitational hair" merges with Kerr-AdS at the OC of gravitational superradiance. In \cite{blackres} these single KVF BHs were constructed numerically, these "black resonators" are periodic and single out a particular frequency. Rather than discuss this numerical construction we will present an explicit construction carried out in \cite{Santos} which approximates the black resonator and gives its leading order thermodynamic properties. This approximation is illuminating as it gives a heuristic insight into the stability properties of these black resonators. An important takeaway from this discussion is that Kerr-AdS BHs are not the only stationary BHs in Einstein-AdS gravity.
\newline \indent

Recall the fundamental behavior of superradaince in global AdS. The amplitude of a mode $e^{-i\omega T+im\Phi}$ can be increased by scattering off a rotating BH with angular velocity $\Omega_h$ when $\omega<m\Omega_h$. In the setting of asymptotically global AdS the boundary allows reflection and so the process of energy extraction repeats leading to an instability. The modes extract enough energy to backreact. The energy gained by the modes has the effect of decreasing $\Omega_h$, eventually leading to a BH with "lumpy hair" rotating around it. Heuristically speaking lumpy, possibly inhomogeneous clumps, of energy co-rotating with the BH would tend to destroy any axisymmetry formerly present in the system; as you move in the $\Phi$ direction you may encounter a more concentrated region of energy. Furthermore, the same can be said of time symmetry. Thus we do not expect $\partial_T$ or $\partial_\Phi$ to be KVFs of the system. However if we simply follow the same clump of hair around we would not expect the metric to vary; i.e the co-rotating vector field $\partial_T + \Omega_h \partial_\Phi$ will be a KVF. We see that instability naturally leads to a BH with a single periodic KVF.
\newline \indent
An object fundamental to the evolution of the superradiant instability is the geon. A geon is a lump of light or energy which is dense enough to be gravitationally bound. Geons can be thought of as nonlinear normal modes of AdS and are solutions that contain only a single Killing field. Any gravitational radiation emitted by the geon is balanced by absorption of waves reflected from the AdS
boundary. They are of interest to us because in \cite{Santos} the black resonator was approximated by placing a small Kerr BH "on top" of a geon. We will review this construction. We first introduce and review some properties about geons and Kerr-AdS:
\newline \indent 
Geons have harmonic time dependence $e^{-i\omega T + i m \Phi}$ in which the centrifugal force balances gravitational attraction. They are horizon free, nonsingular and asymptotically globally AdS. Geons are specified by $l$ and $m$; the number of zeros of the solution along the polar direction and the azimuthal quantum number. They are a 1-parameter family of solutions parameterized by the frequency. At linear order, a geon is a small perturbation around a global AdS background. The energy and angular momentum of the geon are related by $E_g = \frac{\omega}{m}J_g + \mathcal{O}(J^2_g)$; they have zero entropy and an undefined temperature. Note the first law of thermodynamics is obeyed; $dE_g = \frac{\omega}{m} dJ_g$
\newline \indent
For a Kerr-AdS BH with small $E$ and $J$ the leading and next-to leading order thermodynamic quantities are given by: $$E_K \approxeq \frac{r_+}{2}\bigg(1+\frac{r_+^2}{L^2}(1+\Omega_h^2L^2) \bigg) + \mathcal{O}(r_+^4/L^4),~~ J_K \approxeq \frac{1}{2}r_+^3\Omega_h + \mathcal{O}\bigg(\frac{r_+^4}{L^4}\bigg)$$ 
\begin{equation} \label{eq:ther}
S \approxeq \pi r_+^2(1+\Omega_h^2r_+^2) + \mathcal{O}\bigg(\frac{r_+^5}{L^5}\bigg), ~~T_h \approxeq \frac{1}{4\pi r_+}\bigg(1 +(3-2\Omega_h^2L^2)\frac{r_+^2}{L^2}\bigg) +\mathcal{O}\bigg(\frac{r_+^2}{L^2}\bigg)
\end{equation}
Note this also obeys the first law of thermodynamics up to next-to-leading order; $dE_K = \Omega_h dJ_K + T_h dS$.
 The general idea of the construction is that to leading to order the two objects do not interact. The single KVF is inherited from the geon and the charges $E,~J$ of the system are given by $E=E_g+E_K~~\&~~ J=J_g+J_K$
\newline \indent
The entropy and temperature of the final BH are clearly controlled by the Kerr-AdS BH as the geon has zero entropy and undefined temperature. The single KVF chooses the partition of charges between the geon and Kerr-AdS components so as to extremize the total entropy of the system. In particular, maximizing $S=S_K(E-E_g,J-J_g)$ with respect to $J_g$ and considering the first laws for the geon and Kerr-AdS BH, shows that the partition is such that the angular velocities of the two components are the same; $\Omega_h=\frac{\omega}{m}$. Hence the two components are in thermodynamic equilibrium. To see this one can also make the following heuristic argument: Since the geon has only one KVF given by $K=\partial_T + \frac{\omega}{m}\partial_\Phi$ and the Kerr-AdS BH is placed at its center the geon KVF must coincide with the horizon generator of the BH given by $K=\partial_T+\Omega_h\partial_\Phi$. 
\newline \indent
The various restrictions on the system yield the following distribution of charges amongst the components:$$\{J_g,E_g\}=\{J,\frac{\omega}{m}J\},~~\{J_K,E_K\}=\{0,E-\frac{\omega}{m}J\},$$ $$S=4\pi\big(E-\frac{\omega}{m}J\big)^2,~~ T_h=\frac{1}{8\pi}\big(E-\frac{\omega}{m}J\big)^{-1}$$
We see that at leading order the rotation of the system is carried by the geon and the entropy is stored by the Kerr-AdS BH component. These relations obey the first law, $dE=T_hdS+\Omega_hdJ$, up to order $\mathcal{O}(M,J)$, with $\Omega_h = \omega/m$ where $\omega$ is given by eq \ref{eq:ws} for scalar gravitational perturbations and eq \ref{eq:wv} for vector gravitational perturbations.
\newline\indent 
The single KVF BH merges with the Kerr-AdS family at an m-mode onset curve. This occurs when the superradiant condition, $\omega<m\Omega_h$, is saturated. Here $\{\omega,m\}$ are the frequency and azimuthal number of the linearized geon component of the single KVF BH. At the superradaint merger the Kerr-AdS BH and single KVF BH therodynamics coincide, so the Kerr-AdS BH thermodynamics (with $\Omega_h=\omega/m$) can be used to determine the charges of the final system: $$E|_{\text{merg}} \approxeq \frac{r_+}{2} + \frac{r_+^3}{2L^2}\big(1+\frac{\omega^2L^2}{m^2}\big),~~ J|_{\text{merg}} \approxeq \frac{1}{2}\frac{\omega}{m}r_+^3$$
\begin{figure}[h!] 
	\includegraphics[width=16cm]{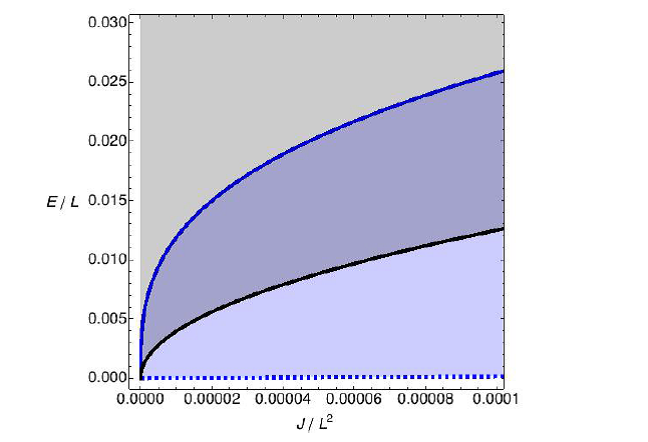}
	\caption{Phase diagram from \cite{Santos} of the stationary solutions of d=4 Einstein-AdS theory. The regions, described in the text, are separated by their stability to the $l=2=m$ scalar modes of gravitational perturbations.}
	\label{fig:fg6}
	\end{figure}
In fig \ref{fig:fg6} we plot a phase diagram $\{E,J\}$ for the $l=2=m$ perturbation mode, with the above curve $\{E_{\text{merg}},J_{\text{merg}}\}$ determining the upper bound of the region where single KVF BHs exist. In the light gray region we have stable Kerr-AdS BH and hence no single KVF BHs. The blue and black curves signify the onset of instability and extremality respectively for Kerr-AdS BH. Hence in the light blue region we have only single KVF BHs. The dashed curve represents the scalar $l=2=m$ geon described by $E=\frac{\omega}{m}J$ with $\omega = \omega_s =1+l$. We see that the black resonators bridge the gap between the onset of superradiance and the geons. In the middle dark gray/blue region we have both Kerr-AdS BH and single KVF BHs; i.e there exists BH pairings (a Kerr BH and a hairy BH) with the same masses and angular momenta but different entropies. Here we have only considered the $l=2=m$ mode, but similar behavior is expected for higher $\{l,m\}$ so we actually have a countably infinite number of examples of non uniqueness for rotating BHs in AdS! 
 
 In \cite{blackres} numerics were used to compare the entropies of a Kerr-AdS BH and the corresponding  black resonator with the same asymptotic charges $E$ and $J$. 
 \begin{figure}[h!] 
 \includegraphics[width = 12cm]{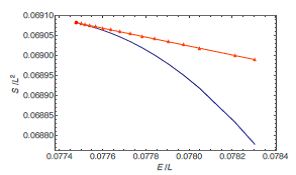}
 \centering
 \caption{Plot from \cite{blackres} comparing the entropy of the Kerr-AdS (blue) and a black resonator (red) of equal asymptotic charges. The leftmost point represents the onset of superradiance. We see that after the onset, for fixed $J$, as the energy is increased the black resonator has the greater entropy of the two}
 \label{fig:ent}
 \end{figure}
 We see from fig \ref{fig:ent} that of the two, the black resonator is the most entropically favorable. 

\subsection{Endpoint of the Superradiant Instability}
The superradiant instability of Kerr-AdS naturally motivates the question of what the end state of the perturbed Kerr BH in AdS is.
We have seen that at the onset of superradiance the Kerr BH merges with a family of single KVF BHs, of which the so called black resonator is one type. These single KVF BHs are certainly a possible intermediate state in the evolution but could they represent the end state? The answer to this question is no. We have seen that any stable candidate for the endpoint must satisfy $\Omega_H L\leq 1$ but it was found numerically that $\Omega_H L >1$ for the black resonators constructed in \cite{blackres}. It is not difficult to see that a single KVF BH associated with a given mode $m$ can only be metastable. Note that while the single KVF is stable to a particular mode $m$ it is not stable to other superradaint modes $m^\prime>m$ which are excited in time evolution because of the nonlinearities in the Einstein equation. To convince yourself of this recall that the black resonator is approximated by a Kerr BH placed inside of a geon. This Kerr BH is the problem because while it will be stable to $m$ as we have seen it will not generally be stable to $m^\prime$. \newline \indent So what is the end state then? At this point we simply don't know. As just mentioned, typically a BH which is stable to perturbation modes $m$ is unstable to modes $m^\prime >m$ so one logically permissible possibility is that system just continues to evolve to black resonators of higher and higher order m. This idea was explored in \cite{coscen} and it was shown that the $m \to \infty$ limit of the black resonator is not a possible end state. Such a solution saturates the bound $E \geq \frac{J}{L}$, required of generic asymptotically AdS solutions with energy-momentum tensor adhering to the dominant energy condition. For details see \cite{coscen}, but it can be shown that this bound is saturated iff the solution is supersymmetric (i.e admits a Killing spinor). The authors were then able to prove that the only supersymmetric vacuum solution which is asymptotically AdS is global AdS itself.
It seems, at present, there are no viable candidates for the end point of the superradiant instability. This leaves two possible outcomes: a singular solution is settled upon in finite time or the system never settles down to a solution. The former violates cosmic censorship as it admits a naked singularity. While for the latter, the development of smaller and smaller structure is driven by the entropically favorable evolution through higher and higher order $m$ black resonators. The ever decreasing scale means that at some point quantum gravitational effects need to be considered. This may be interpreted  as being at odds with cosmic censorship, at least in spirit, because initial data which is well-described classically evolves to a system requiring a quantum mechanical description.

\end{document}